\documentclass[aps,prl,reprint,superscriptaddress,longbibliography]{revtex4-2}

\usepackage{graphicx}
\graphicspath{{./}{./figure/}}

\usepackage{amsmath,amssymb,bm,braket}
\usepackage{enumitem}
\usepackage{comment}

\usepackage{color}

\newcommand{\OLocal}{O}

\usepackage[normalem]{ulem}

\usepackage{hyperref}
\hypersetup{colorlinks=true, citecolor=magenta, linkcolor=blue, urlcolor=cyan}

\usepackage{orcidlink}

\usepackage{amsthm}
\newtheorem{theorem}{Theorem}
\newtheorem{proposition}{Proposition}

\newtheorem{lemma}{Lemma}

\begin{document}
\title{Nonergodicity from Hilbert-Subspace Cloning}

\author{Yuuya Chiba\,\orcidlink{0009-0007-9064-7981}}
\email{yuuya.chiba@phys.s.u-tokyo.ac.jp}
\affiliation{Department of Physics, The University of Tokyo, 7-3-1 Hongo, Tokyo 113-0033, Japan}
\affiliation{Nonequilibrium Quantum Statistical Mechanics RIKEN Hakubi Research Team, Pioneering Research Institute (PRI), RIKEN, 2-1 Hirosawa, Wako, Saitama 351-0198, Japan}

\author{Ryusuke Hamazaki\,\orcidlink{0000-0003-3793-6016}}
\email{ryusuke.hamazaki@riken.jp}
\affiliation{Nonequilibrium Quantum Statistical Mechanics RIKEN Hakubi Research Team, Pioneering Research Institute (PRI), RIKEN, 2-1 Hirosawa, Wako, Saitama 351-0198, Japan}
\affiliation{RIKEN Center for Interdisciplinary Theoretical and Mathematical Sciences (iTHEMS), RIKEN, Wako, Saitama 351-0198, Japan}

\date{\today}

\begin{abstract}
We identify a class of nonintegrable quantum many-body systems composed of single Pauli strings that possess an extensive number of noncommuting local conserved quantities and exhibit partial ergodicity breaking due to exponentially degenerate energy spectra.
We show that such systems exhibit a novel mechanism, \textit{Hilbert-subspace cloning}, where the identical chaotic Hamiltonian acts on each of exponentially many disconnected Hilbert subspaces.
Although the dynamics within each clone thermalizes, quantum coherence between clones remains protected and detectable by local observables. Consequently, the long-time behavior of local observables is described by a coherence-preserving partial Gibbs ensemble, where the hidden subsystem retains its initial quantum coherence.
\end{abstract}

\maketitle

\paragraph{Introduction---}
How an isolated quantum system thermalizes has been a long-standing problem since von Neumann's seminal work about a century ago~\cite{Neumann1929}. Last few decades have witnessed rapid advancement in this field, thanks to the observation of such thermalization dynamics in artificial quantum systems~\cite{Kinoshita2006,Gring2012,Trotzky2012,Kaufman2016,Bernien2017}. While generic quantum many-body systems are believed to locally relax to the thermal state via the eigenstate thermalization hypothesis (ETH)~\cite{Neumann1929,Deutsch1991,Srednicki1994,Rigol2008}, in integrable systems, an extensive number of local conserved quantities prevent the system from thermalizing~\cite{Rigol2007,Rigol2008}. In the last decade, researchers also identified more exotic scenarios where the ergodicity breaks down, such as the quantum many-body scarring~\cite{Bernien2017,Turner2018,Turner2018a,Moudgalya2018,Moudgalya2018a,Schecter2019,Mark2020} and the Hilbert-space fragmentation~\cite{Sala2020,Khemani2020,Moudgalya2021,Moudgalya2022,Yoshinaga2022}. The corresponding Hilbert-space structure is understood by the framework of the commutant algebra~\cite{Moudgalya2022}.

Despite a lot of effort, how the degeneracies of the energy spectrum affect the thermalization dynamics have been less explored. Recent results showed that degeneracies due to nonabelian symmetries~\cite{Murthy2023} or symmetries with projective representation~\cite{Onoda2026} can modify the ETH and lead to anomalous thermalization. It was also found that many degeneracies appear at specific energies as the many-body flat band, leading to nonergodicity~\cite{Jonay2026,Ben-Ami2026}.
However, it remains unclear whether exponentially large spectral degeneracies throughout an otherwise chaotic many-body spectrum can protect quantum coherence, thereby preventing local thermalization.

\begin{figure}
    \centering
    \includegraphics[width=\linewidth]{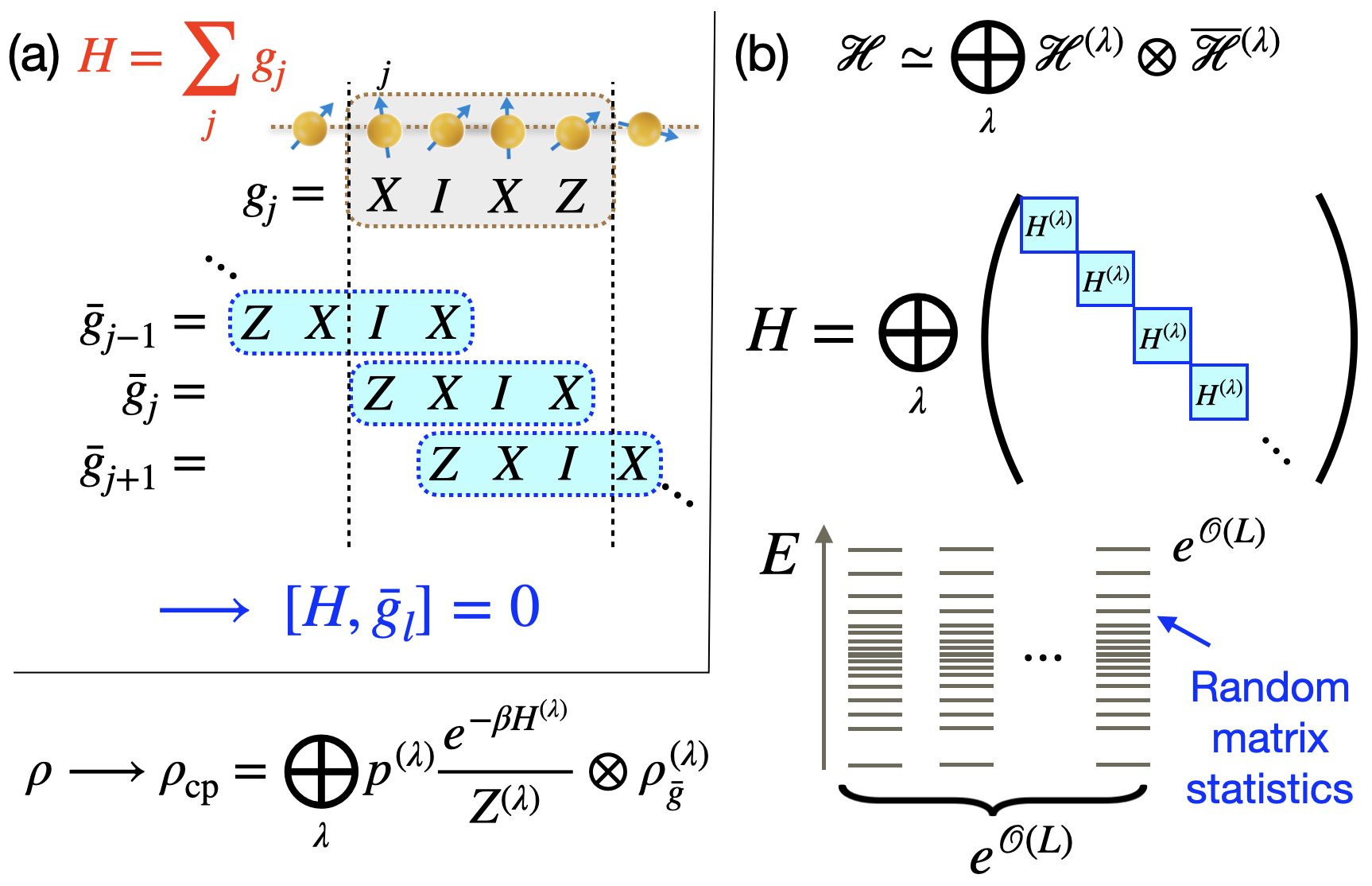}
    \caption{\label{fig:Schematic}
    Schematic illustration of our work. (a) Single-Pauli-string (SPS) Hamiltonian with an extensive number of noncommuting local conserved quantities. Each of the single Pauli operators $g_j$ in the Hamiltonian commutes with mutually noncommuting local operators $\bar g_l$ at each site.
    (b) Hilbert-subspace cloning, which occurs at each global-symmetry sector specified by $\lambda$. The Hamiltonian has a block diagonalized structure, where exponentially many number of identical chaotic Hamiltonians $H^{(\lambda)}$ act on the Hilbert subspaces, each of which has an exponentially large dimension.
    The energy eigenvalue $E$ at each sector then has an exponentially large degeneracies, up to which the spectrum exhibits a chaotic behavior, such as the random-matrix-type level-spacing distribution. 
    The long-time steady state is described by a coherence-preserving partial Gibbs ensemble $\rho_\mathrm{cp}$.
    }
\end{figure}

In this Letter, we propose a novel mechanism of partial nonergodicity due to exponentially large spectral degeneracies in
spin systems hosting an extensive number of local noncommuting conserved quantities.
 These conserved quantities give rise to the \textit{Hibert-subspace cloning (HSC)}, where the identical Hamiltonian satisfying the ETH acts on each of exponentially many disconnected Hilbert subspaces (Fig.~\ref{fig:Schematic}).
 Due to the partial thermalization within each clone and the protection of quantum coherence across clones, the long-time behavior of local observables is described by a coherence-preserving partial Gibbs ensemble, where the  initial quantum coherence is retained in the hidden subsystem.
We demonstrate these scenarios of the HSC by introducing unique single-Pauli-string (SPS) models, where the extensive number of  local noncommuting conserved quantities are naturally obtained by the space-inversion procedure. Our results open a new avenue for the study of nonergodicity in nonintegrable quantum many-body systems.

\paragraph{Single-Pauli-string (SPS) model---}
We start with introducing the concrete model of our interest, which we call the SPS model, as a prototypical model exhibiting the HSC.
We focus on quantum spin-$1/2$ systems on a one-dimensional lattice, described by the Hilbert space $\mathcal{H}=(\mathbb{C}^2)^{\otimes L}$.
Let 
$X_j$, $Y_j$, and $Z_j$
be the Pauli matrices acting on the $j$th site. 
Then, we define the SPS Hamiltonian
as the Hamiltonian written by a sum of a single Pauli string and its translations:
\begin{align}
    H&=\sum_{j=1}^{L}b_j g_{j},\qquad g_{j}=A^{1}_{j}A^{2}_{j+1}...A^{r}_{j+r-1},
    \label{eq:H_Single}
\end{align}
where $b_j\in\mathbb{R}$ can be $j$-dependent, $A^{1},A^{r}\in\{X,Y,Z\}$, $A^{\mu}\in\{I,X,Y,Z\}$ ($\mu=2,3,...,r-1$), and
the periodic boundary condition $A^{\mu}_{j+L}=A^{\mu}_{j}$ is imposed.
Here, $r$ represents the locality of the Hamiltonian;  we utilize the term ``local'' in the sense of short range throughout the manuscript.
Since 
all models with $r\le 3$ become free-fermion solvable~\footnote{For instance, $h_{j}=X_{j}Z_{j+1}Y_{j+2}$ becomes quadratic fermion through Jordan-Wigner transform, and $h_{j}=X_{j}Z_{j+1}Z_{j+2}$ is known as free fermion in disguise~\cite{Fendley2019}.}, the following example with $r=4$~\footnote{From the framework for free-fermion solvability given in Ref.~\cite{Chapman2023}, free-fermion unsolvable cases with $r=4$ are restricted to $g_j=X_jI_{j+1}X_{j+2}Z_{j+3}$, $X_jI_{j+1}Z_{j+2}Z_{j+3}$, $X_jX_{j+1}Z_{j+2}Z_{j+3}$, $X_jZ_{j+1}X_{j+2}Z_{j+3}$ and their equivalents obtained by appropriate permutations of $X,Y,Z$ and their spatial inversions. Note that the third (resp. fourth) case can be written by using a smaller Pauli string with $r=3$ (resp. $r=2$) as $X_jX_{j+1}Z_{j+2}Z_{j+3}=(X_jI_{j+1}Z_{j+2})(X_{j+1}I_{j+2}Z_{j+3})$ [resp. $X_jZ_{j+1}X_{j+2}Z_{j+3}=(X_jZ_{j+1})(X_{j+2}Z_{j+3})$]. In this Letter, we choose the first one rather than the others just for the simplicity of explanation.} serves as the simplest model that ensures nonintegrability (see the discussion later):
\begin{align}
    g_{j}&=X_{j}I_{j+1}X_{j+2}Z_{j+3}.\label{eq:g_XIXZ}
\end{align}

\paragraph{Noncommutative local conserved quantities---}
Now, we describe our first result without restricting it to the above example~\eqref{eq:g_XIXZ}.
We consider the subgroup of the Pauli group that is generated by $g_j$
\begin{align}
    &\mathcal{A}
    =\{c(g_{1})^{n_1}(g_{2})^{n_2}...(g_{L})^{n_L}|\notag\\
    &\hspace{8ex} c\in\{\pm 1,\pm i\},n_1,...,n_L\in\{0,1\}\}.
\end{align}
We also consider the spatial inversion of $g_{j}$,
\begin{align}
    \bar{g}_{j}=A^{r}_{j}A^{r-1}_{j+1}...A^{1}_{j+r-1}.
    \label{eq:gbar}
\end{align}
Crucially, $\bar{g}_{j}$ commutes with $g_{j^{\prime}}$ for all $j,j^\prime$:
\begin{lemma}[\label{lemma:gbar_conservation}Inverted Pauli strings are conserved]
Let $g_j$ be a Pauli string defined in Eq.~\eqref{eq:H_Single} [not necessarily that of Eq.~\eqref{eq:g_XIXZ}], and $\bar{g}_{j}$ be its spatial inversion defined by Eq.~\eqref{eq:gbar}.
Then, they commute with each other
\begin{align}
    [g_j,\bar{g}_{j^{\prime}}]=0
    \label{eq:commutation_g_gbar}
\end{align}
for all $j$ and $j^{\prime}$.
\end{lemma}
\noindent
This is understood from the fact that, due to the spatial inversion structure, there is always an even number of sites at which single-site Pauli operators in $g_j$ and $\bar{g}_{j^{\prime}}$ are anticommuting (see 
Supplemental Material~\cite{SuppMat} for a complete proof).

This also shows that, under the unitary evolution generated by $H$, $\bar{g}_{j}$'s are conserved.
As they are local and noncommutative, $H$ has an extensive number of noncommutative local conserved quantities $\{\bar{g}_{j}\}_{j}$.
Because all terms commute for the Hamiltonian $H$ in the inversion-symmetric case due to $\bar{g}_{j}= g_{j}$, we consider only the inversion-asymmetric case, $\bar{g}_{j}\neq g_{j}$.

\paragraph{Hilbert-subspace cloning---}
Now, we investigate the constraints on the structure of $H$ arising from the existence of the noncommutative local conserved quantities $\{\bar{g}_{j}\}_j$. 
These conserved quantities generate  a subgroup of the Pauli group,
\begin{align}
    &\bar{\mathcal{A}}
    =\{c(\bar{g}_{1})^{n_1}(\bar{g}_{2})^{n_2}...(\bar{g}_{L})^{n_L}|\notag\\
    &\hspace{8ex}c\in\{\pm 1,\pm i\},n_1,...,n_L\in\{0,1\}\},
\end{align}
which is the spatial inversion of $\mathcal{A}$.
For clarity, we focus on example~\eqref{eq:g_XIXZ} and suppose that $L$ is a multiple of $8$.
In this case, the following Pauli strings generate $\mathcal{A}\cap\bar{\mathcal{A}}$ and serve as global symmetries:
\begin{align}
    U_j&:=\prod_{k=0}^{\frac{L}{4}-1}\bar{g}_{4k+j}=\prod_{k=0}^{\frac{L}{4}-1}g_{4k+j+1}
    \quad (j\in\{1,2,3,4\}),
    \label{eq:Uj}\\
    V_j&:=\prod_{k=0}^{\frac{L}{8}-1}\bar{g}_{8k+j}\bar{g}_{8k+j+2}=\prod_{k=0}^{\frac{L}{8}-1}g_{8k+j+5}g_{8k+j+7}\notag\\
    &\hspace{16ex} (j\in\{1,2,...,8\}).
    \label{eq:Vj}
\end{align}
Note that among these elements, only $6$ elements are independent due to their algebraic relation, such as $V_jV_{j+2}=U_{j}$, and hence, we have 
\begin{align}
    \Lambda:=|(\mathcal{A}\cap\bar{\mathcal{A}})/\{\pm1,\pm i\}|=2^6.
\end{align}
Since (elements of) both $\mathcal{A}$ and $\bar{\mathcal{A}}$ commute with $\mathcal{A}\cap\bar{\mathcal{A}}$, the simultaneous eigenspaces of the elements of $\mathcal{A}\cap\bar{\mathcal{A}}$ are invariant under the action of both $\mathcal{A}$ and $\bar{\mathcal{A}}$.
Then, the Hilbert space can be decomposed into these eigenspaces as follows~\cite{SuppMat}:
\begin{theorem}[\label{theorem:Cloning}HSC in the SPS model]
Suppose that $H$ is given by Eqs.~\eqref{eq:H_Single} and \eqref{eq:g_XIXZ}.
There is a decomposition of the Hilbert space,
\begin{align}
    &\mathcal{H}\cong\bigoplus_{\lambda=1}^{\Lambda}\mathcal{H}^{(\lambda)}_{g}\otimes \mathcal{H}^{(\lambda)}_{\bar{g}},
    \label{eq:TheoremCloning_HilbertSpace_Block}
\end{align}
where $\dim\mathcal{H}^{(\lambda)}_{g}=\dim\mathcal{H}^{(\lambda)}_{\bar{g}}=2^{L/2-3}=:d$, 
such that, in a suitable basis,
\begin{align}
    &H=\bigoplus_{\lambda=1}^{\Lambda}H^{(\lambda)}\otimes \bm{1}_d.
    \label{eq:TheoremCloning_HoI}
\end{align}
Here, $H^{(\lambda)}$ acts on $\mathcal{H}^{(\lambda)}_{g}$, and $\bm{1}_d$ is the $d$-dimensional identity.
The decomposition and the choice of the basis are independent of the coefficients $(b_j)_j$.
\end{theorem}
\noindent
In other words, in an appropriate basis, the Hamiltonian within a fixed global-symmetry sector specified by $\lambda$ can be represented as a block-diagonal matrix consisting of exponentially many identical blocks $H^{(\lambda)}$, as shown in Fig.~\ref{fig:Schematic} (b). We call such an exponentially large number of identical-block structure the \emph{Hilbert-subspace cloning (HSC)}. 
This structure guarantees that every energy level has exponentially many degeneracies. 
In our model, this structure and these degeneracies arise from the extensive number of noncommutative conserved quantities, $\bar{g}_j$.

\paragraph{Nonintegrability---}
From Lemma~\ref{lemma:gbar_conservation} and Theorem~\ref{theorem:Cloning}, the model defined by Eqs.~\eqref{eq:H_Single} and \eqref{eq:g_XIXZ} has an extensive number of local conserved quantities $\bar{g}_j$ and exponentially large degeneracies. 
Still, this model has no local conserved quantities independent of $\bar{g}_j$ and satisfies indicators of quantum chaos, suggesting that this model is nonintegrable.

In fact, we can first rigorously show the following:
\begin{theorem}[\label{theorem:XIXZ_LOCQ}Absence of nontrivial local conserved quantities]
Suppose that $H$ is defined by Eqs.~\eqref{eq:H_Single} and \eqref{eq:g_XIXZ} with $b_j\neq 0$ for all $j$.
Then, there are
no local conserved quantities 
except for 
linear combinations of $H$ ($2$-local one) and 
polynomials of $\bar{g}_{j}$'s.
\end{theorem}
\noindent 
See 
Supplemental Material for the precise definition of local conserved quantities and the proof~\cite{SuppMat}.

Next, we investigate the level spacing statistics, one of the indicators of quantum chaos~\cite{Mehta2004,Atas2013,Mondaini2016}.
As Theorem~\ref{theorem:Cloning} shows that every energy level of $H$ has at least $d$ degeneracies, we define the level spacings excluding these $d$ degeneracies. By focusing on each global-symmetry sector specified by $\lambda$ given in Eq.~\eqref{eq:TheoremCloning_HilbertSpace_Block}, the spacing of every $d$ levels of $H$ corresponds to the (nearest-neighbor) level spacing of $H^{(\lambda)}$ defined in Eq.~\eqref{eq:TheoremCloning_HoI}.

\begin{figure}
    \centering
    \includegraphics[width=0.99\linewidth]{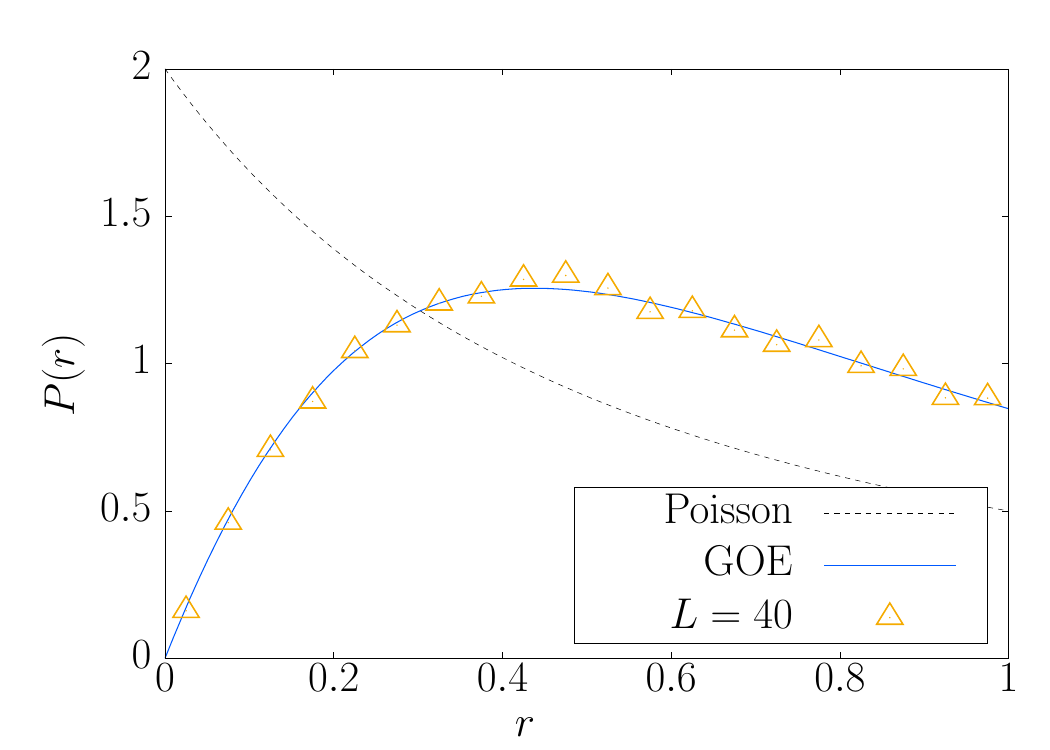}
    \caption{\label{fig:LevelSpacing}Distribution of the ratio of consecutive level spacings of $H^{(\lambda)}$ (orange triangles), which shows the chaotic behavior. The coupling constants $b_j$'s defined in Eq.~\eqref{eq:H_Single} are taken as nonzero irrational numbers that differ for each $j$ (see~\cite{SuppMat} for the detailed parameters). The global-symmetry sector $\lambda$ is chosen such that the eigenvalues of $(V_2,U_3,U_1,V_5,U_2,U_4)$ are $(-1,+1,+1,-1,+1,-1)$. The solid and dashed lines represent the GOE and Poisson distributions, respectively. }
\end{figure}

In Fig.~\ref{fig:LevelSpacing}, we plot the distribution of the ratio of the consecutive level spacings of $H^{(\lambda)}$ (orange triangles) with $L=40$.
The solid (resp. dashed) line depicts the GOE (resp. Poisson) distribution defined by Eq.~(8) [resp. below Eq.~(4)] in Ref.~\cite{Atas2013}. This shows that the level-spacing statistics is well described by the GOE distribution, indicating the quantum chaoticity of $H^{(\lambda)}$.

\paragraph{Eigenstate thermalization hypothesis (ETH)---}
Next, we test whether each subspace-Hamiltonian $H^{(\lambda)}$ satisfies ETH, which is another indicator for quantum chaos.
To this end, we calculate expectation values of physically relevant observables in eigenstates of $H^{(\lambda)}$, denoted by $|\nu\rangle^{\lambda}\in\mathcal{H}_{g}^{(\lambda)}$.
We first focus on local observables acting on the total Hilbert space $\mathcal{H}$, and then derive the corresponding observables that act on the subspace $\mathcal{H}_{g}^{(\lambda)}$.

Without loss of generality, we can restrict our attention to local Pauli strings, as they form a basis of the operator space.
As described in Lemma~\ref{lemma:Generator_g_gbar_X} in End Matter, an arbitrary Pauli string $P$ can be written as
\begin{align}
    P=P_X P_gP_{\bar{g}}.
\end{align}
Here, $P_X$ is a product of Pauli-X operators that connect different global-symmetry sectors,
while $P_g$ (resp. $P_{\bar{g}}$) is a Pauli string generated by using only $g_{j}$'s (resp. $\bar{g}_{j}$'s), which is block-diagonal with respect to $\lambda$.
Since we are interested in the expectation value of the corresponding observable in the eigenstates of $H^{(\lambda)}$, which should vanish when $P_X\neq I$, we consider the $P_X=I$ case. In this case, $P$ can be written as
\begin{align}
    P=P_{g}P_{\bar{g}}=\bigoplus_{\lambda=1}^{\Lambda}P_{g}^{(\lambda)}\otimes P_{\bar{g}}^{(\lambda)},
    \label{eq:ETH_P=oplus_lambda}
\end{align}
where $P_{g}^{(\lambda)}$ [resp. $P_{\bar{g}}^{(\lambda)}$] is a matrix on $\mathcal{H}_{g}^{(\lambda)}$ [resp. $\mathcal{H}_{\bar{g}}^{(\lambda)}$] corresponding to $P_{g}$ [resp. $P_{\bar{g}}$].
In summary, we test the ETH that states that the expectation values ${}^{\lambda}\langle\nu|P_{g}^{(\lambda)}|\nu\rangle^{\lambda}$ are sufficiently close to a smooth function independent of $\nu$ for the case where $P$ is a local Pauli string.

\begin{figure}
    \centering
    \includegraphics[width=0.99\linewidth]{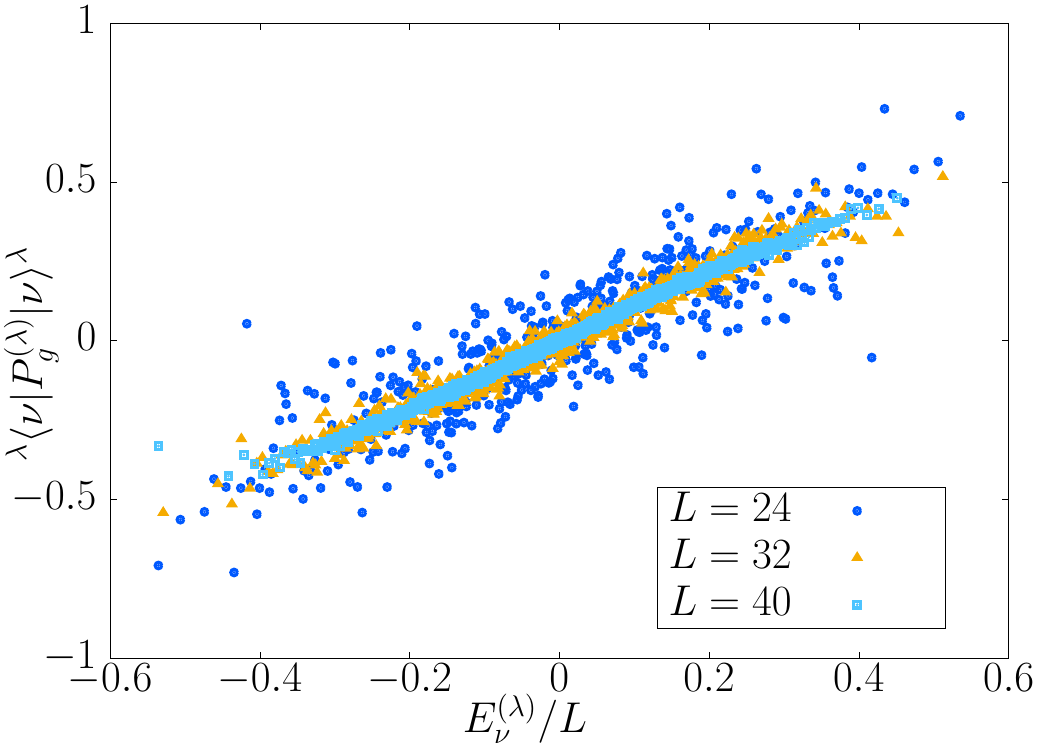}
    \caption{\label{fig:ETH}Expectation value of a matrix $P_{g}^{(\lambda)}$ on $\mathcal{H}_{g}^{(\lambda)}$ corresponding to a Pauli string $P=g_{1}$ in eigenstates $|\nu\rangle^{\lambda}$ of $H^{(\lambda)}$ with eigenenergy $E_{\nu}^{(\lambda)}$. The coupling constants $b_j$'s defined in Eq.~\eqref{eq:H_Single} are taken as the same as in Fig.~\ref{fig:LevelSpacing}. The global-symmetry sector $\lambda$ is chosen such that the eigenvalues of $(V_2,U_3,U_1,V_5,U_2,U_4)$ are $(-1,+1,+1,+1,+1,+1)$.}
\end{figure}

In Fig.~\ref{fig:ETH}, we plot ${}^{\lambda}\langle\nu|P_{g}^{(\lambda)}|\nu\rangle^{\lambda}$ for each eigenstate $|\nu\rangle^{\lambda}$ of $H^{(\lambda)}$ with eigenenergy $E_{\nu}^{(\lambda)}$.
Here, the Pauli string $P$ that we focus on is taken as $P=g_{1}$. As $L$ is increased from $L=24$ to $40$, all points at the bulk of energy spectrum approach a single curve, indicating that $H^{(\lambda)}$ satisfies the ETH.

\paragraph{Nonergodicity and anomalous thermalization---}
While the ETH is numerically verified above, our model shows anomalous relaxation due to the exponentially large degeneracy.
Under physically reasonable assumptions, we obtain the expression for the state after relaxation, which preserves coherence between degenerate energy eigenstates (see 
Supplemental Material~\cite{SuppMat} for a proof and a more formal version):
\begin{theorem}[\label{theorem:Thermalization_Informal}Thermalization to coherence-preserving partial Gibbs state]
Let $H$ be the Hamiltonian defined by Eqs.~\eqref{eq:H_Single} and \eqref{eq:g_XIXZ}.
Take $b_j$'s nonuniform and nonzero values, breaking the translation symmetry, so that $H$ has no degeneracy other than $d$-fold degeneracy predicted by Theorem~\ref{theorem:Cloning}.
Suppose that the initial state $\rho$ has a macroscopically definite energy density corresponding to a finite inverse temperature $\beta$.
Take a local observable $\OLocal$ of interest.
Assume ETH in the sense above, meaning that ${}^{\lambda}\langle\nu|P_{g}^{(\lambda)}|\nu\rangle^{\lambda}$ is sufficiently close to a smooth function independent of $\nu$ for any $P_{g}^{(\lambda)}$ corresponding to a local Pauli string $P$.

Consider a unitary time evolution $\rho(t)=e^{-iHt}\rho e^{iHt}$.
Then, the long-time average of the expectation value of $\OLocal$,
\begin{align}
    \overline{\mathrm{Tr}[\rho(t) \OLocal]}=\lim_{\mathcal{T}\to\infty}\frac{1}{\mathcal{T}}\int_{0}^{\mathcal{T}} dt\ \mathrm{Tr}[\rho(t) \OLocal],
\end{align}
agrees with its expectation value in 
\begin{align}
    \rho_{\mathrm{cp}}=\bigoplus_{\lambda=1}^{\Lambda}p^{(\lambda)}\frac{e^{-\beta H^{(\lambda)}}}{Z^{(\lambda)}}\otimes \rho^{(\lambda)}_{\bar{g}},
    \label{eq:rho_cp}
\end{align}
where
\begin{align}
    p^{(\lambda)}&:=\mathrm{Tr}[\rho\Pi^{(\lambda)}],\\
    \rho^{(\lambda)}_{\bar{g}}&:=\mathrm{Tr}_{g}^{(\lambda)}[\Pi^{(\lambda)}\rho\Pi^{(\lambda)}]/p^{(\lambda)}.
\end{align}
Here, $\Pi^{(\lambda)}$ is the projection operator onto the global-symmetry sector $\lambda$, and  $\mathrm{Tr}_{g}^{(\lambda)}$ is the partial trace on that sector over $\mathcal{H}^{(\lambda)}_{g}$.

\end{theorem}

\noindent
This means that, in our model, $\rho$ relaxes to the state $\rho_{\mathrm{cp}}$, which preserves the initial coherence between degenerate energy eigenstates, represented by $\rho^{(\lambda)}_{\bar{g}}$. We call $\rho_{\mathrm{cp}}$ the \emph{coherence-preserving partial Gibbs state}.
Since this state differs from the ordinary Gibbs state $e^{-\beta H}/Z$, relaxation to the ordinary Gibbs state, namely thermalization, does not occur.
This means that the time average and the (ordinary) ensemble average do not agree, indicating \emph{nonergodicity}.

We have more comment on Theorem~\ref{theorem:Thermalization_Informal}.
The equilibrium values of thermodynamically relevant observables are usually independent of the choice of the global-symmetry sectors $\lambda$.
This suggests $\frac{e^{-\beta H^{(\lambda)}}}{Z^{(\lambda)}}\simeq \frac{e^{-\beta H^{(\lambda=1)}}}{Z^{(\lambda=1)}}$, and Eq.~\eqref{eq:rho_cp} reduces to 
\begin{align}
    \frac{e^{-\beta H^{(\lambda=1)}}}{Z^{(\lambda=1)}}\otimes \Bigl(\bigoplus_{\lambda=1}^{\Lambda}p^{(\lambda)}\rho^{(\lambda)}_{\bar{g}}\Bigr).
    \label{eq:rho_cp_approx}
\end{align}
On the other hand, one might naively suspect that, by identifying $\mathcal{H}_{\bar{g}}^{(\lambda)}$'s with a common Hilbert space $\mathcal{H}_{\bar{g}}\cong\mathbb{C}^d$ and writing the Hamiltonian in a tensor product form,
\begin{align}
    &\mathcal{H}\cong\tilde{\mathcal{H}}\otimes \mathcal{H}_{\bar{g}},\quad \tilde{\mathcal{H}}:=\bigoplus_{\lambda=1}^{\Lambda}\mathcal{H}^{(\lambda)}_{g},\\
    &H=\tilde{H}\otimes \bm{1}_d,\quad \tilde{H}:=\bigoplus_{\lambda=1}^{\Lambda}H^{(\lambda)},
\end{align}
the state after relaxation could be described by
\begin{align}
    \frac{e^{-\beta \tilde{H}}}{\tilde{Z}}\otimes \mathrm{Tr}_{\tilde{\mathcal{H}}}[\rho].
    \label{eq:WrongSteadyState}
\end{align}
However, this is not the case because the $\lambda $-dependent coherence in $\rho^{(\lambda)}_{\bar{g}}$, which can be detected by local observables as illustrated in an example
given in End Matter, is lost by averaging over all $\lambda$,
\begin{align}
    \mathrm{Tr}_{\tilde{\mathcal{H}}}[\rho]=\sum_{\lambda=1}^{\Lambda}p^{(\lambda)}\rho^{(\lambda)}_{\bar{g}},
\end{align}
which should be contrasted with Eqs.~\eqref{eq:rho_cp} and \eqref{eq:rho_cp_approx}.
This physically relevant dependence on the global-symmetry sectors $\lambda$ when describing the state after relaxation
is a remarkable feature of HSC that ordinary thermalizing systems do not exhibit.

\paragraph{Hidden subsystem retaining coherence---}

We can take a stabilizer group generated by
\begin{align}
    \{g_{2k+1},\bar{g}_{2k+1}\}_{k=0}^{L/2-1}.
\end{align}
Through conjugation by a suitable Clifford gate, we can map these Pauli strings to the Pauli-$Z$ operators $Z_j$'s as follows (see~\cite{SuppMat} for a proof and a formal statement):
\begin{proposition}[\label{proposition:HiddenSubsystem}Clifford gate revealing a hidden half-length subsystem]
There is a Clifford gate $C$ that induces the following transformation through conjugation $C\bullet C^{\dagger}$:
\begin{align}
    \bar{g}_{2k+1}&\mapsto Z_{2k+1}\\
    g_{2k+1}&\mapsto Z_{2k+2}
    \\
    \bar{g}_{2k+2}&\mapsto X_{2k-1}X_{2k+1}X_{2k+3}X_{2k+5}\\
    g_{2k+2}&\mapsto X_{2k}X_{2k+2}X_{2k+4}X_{2k+6}
    \label{eq:Mapping_XXXX}
\end{align}
except for $\bar{g}_{L-4}$, $\bar{g}_{L-2}$, $\bar{g}_{L}$, $g_{L-4}$, $g_{L-2}$, $g_{L}$, 
on which $\lambda$-dependent boundary conditions are imposed.
Neglecting such boundary corrections, we can write the Hamiltonian defined by Eqs.~\eqref{eq:H_Single} and \eqref{eq:g_XIXZ} after the transformation as
\begin{align}
    H&\mapsto\sum_{k=0}^{L/2-1}b_{2k+1}Z_{2k+2}+\sum_{k=0}^{L/2-1}b_{2k+2}X_{2k}X_{2k+2}X_{2k+4}X_{2k+6}.
    \label{eq:H_AfterClifford}
\end{align}
\end{proposition}
\noindent
The Hamiltonian~\eqref{eq:H_AfterClifford} is defined only on the even sites, indicating $2^{L/2}$ degeneracy (precisely, the boundary corrections explain the exact value of the degeneracy $2^{L/2-3}$).
Furthermore, the nonintegrability proved in Theorem~\ref{theorem:XIXZ_LOCQ} and chaoticity numerically verified in Figs.~\ref{fig:LevelSpacing} and \ref{fig:ETH}
could be understood as the corresponding properties of the Hamiltonian~\eqref{eq:H_AfterClifford}.
This would help us understand the phenomena observed in our model more easily.

\paragraph{Extension---}
Finally, we discuss extensions of our results. 
The anomalous behaviors of our model stem from the structure given by Theorem~\ref{theorem:Cloning}, which has a broader scope of application than presented in this Letter. For instance, it applies not only to $H$ written as a linear combination of $g_j$'s as in Eq.~\eqref{eq:H_Single}, but also a polynomial of $g_j$'s, e.g., $H=\sum_j(g_j+g_jg_{j+1}+g_jg_{j+2})$.
Furthermore, it also applies to a circuit that can be generated by $g_j$'s; for example, we can consider a Floquet circuit composed of $U=\prod_{k=1}^{L/4} g_{4k} \prod_{k=1}^{L/4} g_{4k-1} \prod_{k=1}^{L/4} g_{4k-2} \prod_{k=1}^{L/4} g_{4k-3}$ with Eq.~\eqref{eq:g_XIXZ}, which satisfies $[U,\bar{g}_j]=0$.
The extension to higher-dimensional lattices is also straightforward, provided that the spatial inversion in Lemma~\ref{lemma:gbar_conservation} is replaced by its higher-dimensional counterpart.

\paragraph{Conclusion---}
We have constructed the single-Pauli string (SPS) Hamiltonians that exhibit Hilbert-subspace cloning (HSC), where an expontially  degenerate energy spectrum leads to the nonergodicity despite the system's nonintegrable structure.
Assuming the ETH for each cloned Hamiltonian, we find that the 
steady-state expectation value of local observables is described by a coherence-preserving partial Gibbs state.

Our work opens many directions for future research. One important question is how the protected quantum coherence affects dynamical properties beyond stationary expectation values, such as transport and operator scrambling. Another intriguing direction is to explore the interplay between HSC and other mechanisms of ergodicity breaking. For instance, the dynamics may become even richer when the cloned Hamiltonians host quantum many-body scars. Furthermore, the Hilbert-space structure underlying cloning realizes the noiseless subsystems~\cite{Lidar2014} studied in quantum information theory. This connection suggests potential applications to quantum information storage and manipulation in interacting many-body systems. Finally, the ground-state properties of SPS Hamiltonians also merit further study, particularly in view of their extensive local noncommuting conserved quantities.

\paragraph{Note added---} 
After completion of our work, we became aware of Ref.~\cite{Pushkar2026}, which discusses another nonintegrable model with a large number of degeneracies due to anti-commuting conserved quantities.

\begin{acknowledgments}
We are grateful to H.~Yoshida, S.~Sugimoto, and K.~Fukai for useful discussions.
Generative AI tools (ChatGPT 6) were used to improve the readability of the manuscript. Interactions with these tools helped find the example given in End Matter. All results were independently verified by the authors.
Y.C. is supported by the Special Postdoctoral Researchers Program at RIKEN.
R.H.\ is supported by JSPS KAKENHI Grant No.~JP24K16982.
Y.C. and R.H. are supported by JST ERATO Grant No.~JPMJER2302, Japan.
\end{acknowledgments}

\bibliographystyle{apsrev4-2}
\bibliography{Letter.bbl}

\clearpage
\section*{End Matter}

\subsection*{Structure of energy eigenbasis}

Now, we investigate matrix elements of an arbitrary observable in the energy eigenstate basis. 
To deal with arbitrary observables efficiently, we first show that all Pauli strings can be generated by certain generators of $\mathcal{A}$ and $\bar{\mathcal{A}}$ in addition to six Pauli-X operators $X_1,X_2,...,X_6$.
\begin{lemma}[\label{lemma:Generator_g_gbar_X}Generators of all Pauli strings]
All Pauli strings can be generated by the following elements: $X_j$ ($j\in\{1,2,...,6\}$), 
\begin{align}
    g_{8k+1},\ g_{8k+2}g_{8k+4},&\ g_{8k+3},\ g_{8k+5},\nonumber\\
    \bar{g}_{8k+1},\ \bar{g}_{8k+2}\bar{g}_{8k+4},&\ \bar{g}_{8k+3},\ \bar{g}_{8k+5},\nonumber\\
    &(k\in\{0,1,...,L/8-1\}),
    \label{eq:Stabilizer_g_gbar}
\end{align}
and
\begin{align}
    g_{2},\ 
    g_{8k+4}g_{8k+6},\ 
    g_{8k+7},\ 
    g_{8k+6}g_{8k+8},\ 
    g_{8k+8}g_{8k+10}, \notag\\
    \bar{g}_2,\ 
    \bar{g}_{8k+4}\bar{g}_{8k+6},\ 
    \bar{g}_{8k+7},\ 
    \bar{g}_{8k+6}\bar{g}_{8k+8},\ 
    \bar{g}_{8k+8}\bar{g}_{8k+10}, \notag\\
    (k\in\{0,1,...,L/8-2\}).
    \label{eq:Generator_g_gbar}
\end{align}
Furthermore, since this generating set consists of only $2L$ number of Pauli strings, these operators are independent.
\end{lemma}
\noindent 
This lemma implies that an arbitrary observable can be written as a polynomial of the above generators.
The proof of this lemma is given in Supplemental Material~\cite{SuppMat}.

From Lemma~\ref{lemma:Generator_g_gbar_X}, an arbitrary Pauli string $P$ can be decomposed into three parts uniquely
\begin{align}
    P=P_XP_{g}P_{\bar{g}},
    \label{eq:P1P2P3}
\end{align}
where $P_X$ is a product of $X_j$ ($j\in\{1,2,...,6\}$),
$P_{g}$ is a product of terms in Eqs.~\eqref{eq:Stabilizer_g_gbar} and \eqref{eq:Generator_g_gbar} consisting of $g_j$'s,
and $P_{\bar{g}}$ is a product of terms in Eqs.~\eqref{eq:Stabilizer_g_gbar} and \eqref{eq:Generator_g_gbar} consisting of $\bar{g}_j$'s.

Now, we investigate the matrix elements of $P$ in the energy eigenbasis with degenerate eigenenergy.
Let $\ket{\lambda,\nu,m}$ be an eigenstate of $H$ with eigenenergy $E_{\nu}^{(\lambda)}$ in the global-symmetry sector $\lambda$,
where $\nu\in\{1,2,...,d\}$ is the label of eigenenergy and $m\in\{1,2,...,d\}$ is an appropriately chosen label for degeneracy.
The corresponding eigenvector of $\lambda$th block Hamiltonian $H^{(\lambda)}\otimes \bm{1}_d$ is denoted by $|\nu\rangle^{\lambda}\otimes |m\rangle^{\lambda}$.
Then, we obtain the following (see 
Supplemental Material~\cite{SuppMat}):
\begin{proposition}[\label{proposition:WignerEckart_Informal}Wigner-Eckart type factorization]

Take an arbitrary Pauli string $P$, which can be written as \eqref{eq:P1P2P3}. 
If $P_X\neq \bm{1}_{2^L}$, we have
\begin{align}
    \bra{\lambda,\nu,m^\prime}P\ket{\lambda,\nu,m}=0.
    \label{eq:WignerEckart0_Informal}
\end{align}
Otherwise (i.e., $P_X=\bm{1}_{2^L}$), $P$ can be written as
\begin{align}
    P=P_gP_{\bar{g}}=\bigoplus_{\lambda=1}^{\Lambda}P_{g}^{(\lambda)}\otimes P_{\bar{g}}^{(\lambda)},
\end{align}
and we have
\begin{align}
    &\bra{\lambda,\nu,m^\prime}P\ket{\lambda,\nu,m}\notag\\
    &={}^\lambda\langle\nu|P_g^{(\lambda)}|\nu\rangle^{\lambda}
    {}^{\lambda}\langle m^\prime|P_{\bar{g}}^{(\lambda)}|m\rangle^{\lambda}.
    \label{eq:WignerEckart_Informal}
\end{align}
\end{proposition}
\noindent
Note that Eq.~\eqref{eq:WignerEckart0_Informal} follows from the fact that, for all $P_X\neq I$, the diagonal Blocks with respect to $\lambda$ are zero.
The main claim of Proposition~\ref{proposition:WignerEckart_Informal} is Eq.~\eqref{eq:WignerEckart_Informal}.
It shows that if we know the value of a single element $\bra{\lambda,\nu,m^\prime}P\ket{\lambda,\nu,m}$, then we can obtain ${}^\lambda\langle\nu|P_g^{(\lambda)}|\nu\rangle^{\lambda}$, and hence, all other elements $\{\bra{\lambda,\nu,m^\prime}P\ket{\lambda,\nu,m}\}_{m^\prime,m}$ by calculating simpler elements ${}^{\lambda}\langle m^\prime|P_{\bar{g}}^{(\lambda)}|m\rangle^{\lambda}$.
In addition, the ETH for the Hamiltonian $H^{(\lambda)}$ introduced below Eq.~\eqref{eq:ETH_P=oplus_lambda} corresponds to requiring that ${}^\lambda\langle\nu|P_g^{(\lambda)}|\nu\rangle^{\lambda}$ is close to a sufficiently smooth function.

\subsection*{Example showing that Eq.~(\ref{eq:WrongSteadyState}) is wrong}
Here, we show an example where Eq.~\eqref{eq:WrongSteadyState} is wrong.
Consider a state $\rho=(\bm{1}_{2^L}+\bar{g}_3)/2^L$, and focus on $\OLocal=\bar{g}_3$.
Since $\bar{g}_3$ is conserved, we have $\rho(t)=\rho$, and hence
\begin{align}
    \overline{\mathrm{Tr}[\rho(t) \OLocal]}=1.
\end{align}
Next, we decompose $\rho$ as
\begin{align}
    \rho=\frac{1}{2^L}\bigoplus_{\lambda=1}^{\Lambda}\bm{1}_d\otimes (\bm{1}_d+\bar{g}_3^{(\lambda)}),
\end{align}
where we have used Eq.~\eqref{eq:ETH_P=oplus_lambda}.
From this expression, we have $\beta=0$, $ p^{(\lambda)}=1/\Lambda$, and $\rho^{(\lambda)}_{\bar{g}}=(\bm{1}_d+\bar{g}_3^{(\lambda)})/d$.
Then, the coherence-preserving partial Gibbs state in Eq.~\eqref{eq:rho_cp} is given by
\begin{align}
    \rho_{\mathrm{cp}}=\frac{1}{2^L}\bigoplus_{\lambda=1}^{\Lambda}\bm{1}_d\otimes (\bm{1}_d+\bar{g}_3^{(\lambda)})=\rho,
\end{align}
and hence, it correctly describes the state after relaxation.
On the other hand, as shown in 
Supplemental Material~\cite{SuppMat}, we have 
\begin{align}
    \mathrm{Tr}_{\tilde{\mathcal{H}}}[\rho]=\bm{1}_d/d,
    \label{eq:CancelCoherence}
\end{align}
which implies that $\mathrm{Tr}\bigl[\frac{e^{-\beta \tilde{H}}}{\tilde{Z}}\otimes \mathrm{Tr}_{\tilde{\mathcal{H}}}[\rho]O\bigr]=0$.
This means that the state after relaxation is not described by Eq.~\eqref{eq:WrongSteadyState}.




\end{document}